\documentclass[]{article}

\usepackage{latexsym}      
\usepackage[pdftex]{color, graphicx}      
\usepackage{amsmath}
\usepackage{subfigure}
\usepackage[pdftex]{hyperref}

\providecommand{\bra}[1]{\langle#1\rvert}
\providecommand{\ket}[1]{\lvert#1\rangle}

\title{Optimization of the damped quantum search}
\author{Neris Ilano, Cristine Villagonzalo and Ronald Banzon\\
\vspace{0.015in}
\textit{\footnotesize National Institute of Physics, University of the Philippines, }\\
\textit{\footnotesize Diliman Quezon City, 1101 Philippines}}

\begin{document}

\maketitle

\begin{abstract}
The damped quantum search proposed in [A. Mizel \textit{Phys. Rev. Lett.} \textbf{102} 150501 (2009)] was analyzed by calculating the highest possible probability of finding the target state in each iteration. A new damping parameter that depends on the number of iterations was obtained, this was compared to the critical damping parameter for different values of target to database size ratio. The result shows that the range of the new damping parameter as a function of the target to database size ratio increases as the number of iterations is increased. Furthermore, application of the new damping parameter  per iteration on the damped quantum search scheme shows a significant improvement on some target to database size ratio (i.e. $\geq50\%$ maximum percentage difference)  over the critically damped quantum search.
\end{abstract}

\section{Introduction}
Classically, unsorted database are searched by inspecting each element and checking if they satisfy the desired property. If $N$ is the size of the database and $M$ is the number of target items, then the time complexity of obtaining the one of the desired element is $\mathcal{O}(N/M)$. In \cite{grover0}, Grover introduced a quantum search algorithm which can speed up the process to $\mathcal{O}(\sqrt{N/M})$ operations. A proof that Grover's search algorithm is the best possible oracle based search algorithm was developed in \cite{Bennett, Boyer}. The generality of the search based problem makes Grover's algorithm one of the interesting field of research in quantum computing. Some research are focused on its application \cite{Liu, Aaronson}, while others on its extension \cite{Cafaro, Childs, Farhi}.

One notable property of the quantum search algorithm is its oscillatory nature, i.e. the probability of finding one of the target states oscillates from zero to some maximum value. The number of queries that will give the maximum probability of success is dependent on the number of target states in the database \cite{Nielsen, Brylinsky}. Searching a quantum database using Grover's algorithm without a prior knowledge of the number of target items poses a problem because the probability of finding a target state successfully does not converge as the number of iterations is increased. Several modifications were proposed to address this dilemma \cite{fix1,fix2, Grover2} which, however, cannot preserve the $\mathcal{O}(N/M)$ signature. An alternative way of solving the problem is proposed in \cite{mizel} where the quantum search algorithm is damped by attaching an external spin to the quantum database. In this new scheme, knowledge of the number of target items in the database is not needed.

In the damped quantum search algorithm \cite{mizel} there exists a critical damping parameter which is introduced using a physical argument. The critical damping parameter suppresses the oscillations and the error probability from increasing. By simulating the behavior of the critical damping parameter, the author was able to propose a damping parameter which varies with each iteration. The results were then able to suppress the oscillation of the success probability as the search approaches one of the target states.

In this work, we obtained the optimized damping parameter up to the tenth iteration by calculating the failure probability. This was done by solving the absolute minima of the failure probability for a given number of target states and database size as a function of the damping parameter per iteration. We also examined the critical damping parameter by comparing it with the optimized one. The goal of this work is to analyze if the critical damping is optimized. We will do this by directly probing the behavior of the probability of obtaining the target state as the number of target states is increased.

\section{Damped Quantum Search}
\noindent
Suppose we have a total number of states $N$, out of which $M$ are target states. Let $\ket{\psi}=\cos\theta/2\ket{\alpha}+\sin\theta/2\ket{\beta}$, where $\ket{\alpha}$ is the equal superposition of the non-target states, $\ket{\beta}$ is the equal superposition of the target states and $\sin\theta/2=\sqrt{M/N}$.
The action of the Grover's algorithm is encapsulated in the operator $G$. For a detailed discussion of the Grover's algorithm, we refer the reader to \cite{Nielsen}.

The damped quantum search introduced in \cite{mizel} modified the Grover formulation as follows.
First, an external spin $\ket{\downarrow}$ is appended to the initial state $\ket{\psi}$ (i.e. $\ket{\psi} \longrightarrow \ket{\psi}\otimes\ket{\downarrow}$).  Then, the following operator is introduced
\begin{equation}\label{eq:U}
U=\left[G\frac{1-S_{z}}{2}+\frac{1+S_{z}}{2}\right]
\left[e^{-i\phi S_{y}}\frac{1-Z}{2}+\frac{1+Z}{2}\right]
\end{equation}
where $S_{y}$ and $S_{z}$ are the external Pauli spin matrices acting in the Hilbert space of the externally appended spin, the oracle $Z=\ket{\alpha}\bra{\alpha}-\ket{\beta}\bra{\beta}$ acts on the Hilbert space of the database and $\phi$ is the damping parameter. The action of $U$ is described as follows: (i) the right factor calls the oracle $Z$ and flips the external spin if $\ket{\psi}=\ket{\beta}$, (ii) the left factor utilizes $G$ only if the external spin has not flipped. In this way, the external spin limits the application of $G$ as the target state is reached.

Each application of $U$ is followed by a measurement of the external spin. If the external spin has flipped, then the iteration stops. Without knowing the measurement result, the system composite density matrix assumes the form $(1-\mathbf{Tr}\rho)\ket{\beta}\bra{\beta}\otimes\ket{\uparrow}\bra{\uparrow}
+\rho\otimes\ket{\downarrow}\bra{\downarrow}$ where $\rho$ is initially $\ket{\psi}\bra{\psi}$. This means that the probability that the system will collapse to $\ket{\beta}\bra{\beta}\otimes\ket{\uparrow}\bra{\uparrow}$, i.e. the target state has been found and the external spin has flipped, is $1-\mathbf{Tr}\rho$. Furthermore, each iteration yields the following:
\begin{equation}\label{eq:mat}
\begin{bmatrix}
\mathbf{Tr}(\rho'X) \\
\mathbf{Tr}(\rho'Z) \\
\mathbf{Tr}(\rho')  \\
\end{bmatrix}
=
\begin{bmatrix}
\cos 2\theta\cos\phi  &\sin 2\theta\frac{1+\cos^{2}\phi}{2} &\sin 2\theta\frac{1-\cos^{2}\phi}{2} \\
-\sin 2\theta\cos\phi &\cos 2\theta\frac{1+\cos^{2}\phi}{2} &\cos 2\theta\frac{1-\cos^{2}\phi}{2} \\
0 					  &\frac{1-\cos^{2}\phi}{2}             &\frac{1+\cos^{2}\phi}{2}             \\
\end{bmatrix}
\begin{bmatrix}
\mathbf{Tr}(\rho X) \\
\mathbf{Tr}(\rho Z) \\
\mathbf{Tr}(\rho)   \\	
\end{bmatrix}		
\end{equation}
where $X=\ket{\beta}\bra{\alpha}+\ket{\alpha}\bra{\beta}$ and $[\mathbf{Tr}(\rho X)\;\mathbf{Tr}(\rho Z)\; \mathbf{Tr}(\rho)]^{T}=[\sin\theta\;\cos\theta\;1]^{T}$. The combination of U and the measurement of the external spin increases $1-\mathbf{Tr}\rho$ as the number of iterations is increased. The average number of oracle queries to find the target item for small $\phi$ has the quantum search signature $O(\sqrt{N})$, while for $\phi\sim\pi/2$, it takes an average of $O(N)$ queries, which is the classical search limit. The existence of a critical damping defined by $\cos\phi_{crit}=(1-\sin\theta)/(1+\sin\theta)$ divides the two regimes. When $\phi=\phi_{crit}$, the three eigenvalues of the square matrix in Eq. \ref{eq:mat} are equal.
It turns out that even for the weakest possible value of $\phi$, where $M$ is set to 1 for arbitrary $N$, the damping is still evident. However, this smallest value of $\phi$ can only damp the search effectively if $M<N/2$. In the absence of the knowledge of $M$ and in case $M\geq N/2$, the critical damping varies from iteration to iteration according to $\cos\phi_{r}=(1-\sin(\pi/2r))/(1+\sin(\pi/2r))$, where $r$ corresponds to the $r^{th}$ iteration. This proposed variation of $\phi$ starts with the largest damping and weakens with increasing number of iterations. This variation is designed to mimic the behavior of the critical damping.

\section{Analytical calculation of $\mathbf{Tr}\rho(\theta,\phi)$}
\begin{figure}[!htp]
\centering
\includegraphics[trim = 0mm 0mm 0mm 160mm, clip, height=2in, width=4in]{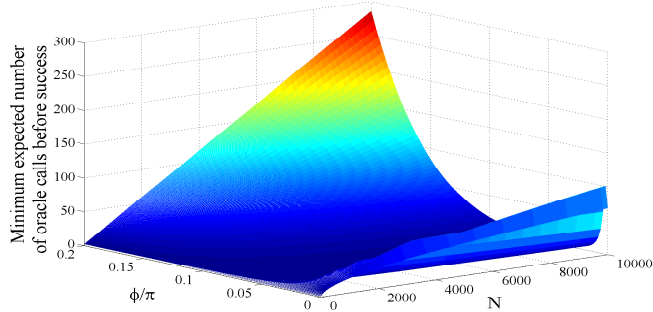}
\caption{\label{fig:Fig2_refined}Expected number of oracle calls before success as a function of the database size, $N$, and damping parameter, $\phi$.}
\end{figure}

\noindent
Given the size of the database $N$ and the damping parameter $\phi$, we execute $n$ calls to $U$. The transition of the external spin from $\ket{\downarrow}$ to $\ket{\uparrow}$ indicates that the target has been found. However, if the external spin remains unaltered then we perform an additional query to the oracle to check if the target has been found. If the outcome turns out to be negative, then we start the procedure all over again. Fig.~\ref{fig:Fig2_refined} shows the minimum expected number of calls before success, $E(n)=(n+1)/P(n)$, where $P(n)$ is the probability of success after the $n^{th}$ call as given by 
\begin{equation}
P(n)\propto \Bigl\lvert\bigl(\bra{\beta}\otimes\bra{\uparrow}
+\bra{\beta}\otimes\bra{\downarrow}\bigr)
\cdot \bigl(U^{n}\ket{\psi}\otimes\ket{\downarrow}\bigr)\Bigr\rvert^{2}.
\end{equation}

The linear behavior in the region of large damping parameter portrays the classical result, while the region of small damping parameter shows the quadratic quantum character. The valley separating these two regimes is interpreted as the region that has the critical damping, $\phi_{crit}$, based from the eigenvalues of the square matrix in Eq.~(\ref{eq:mat}).

We analyze the distinct valley to determine if $\phi_{crit}$ is the same as the damping parameter that gives the smallest possible $\mathbf{Tr}\rho$. We refer to this as the \textit{optimum damping}, $\phi_{opt}$. To start with our analysis, let us define a positive number that serves as an upper bound to the probability that the external spin has not flipped, such that
\begin{equation}
  \mathbf{Tr}\rho\leq\delta.
\end{equation}
We call $\delta$ the \textit{failure tolerance}. 

Suppose we perform $n$ iterations, what is the smallest possible failure tolerance, $\delta_{min}$, that we can assign? What is the value of the damping parameter $(\phi_{opt})$ that will give a $\delta_{min}$? Is the computed $\phi_{opt}$ the same as the $\phi_{crit}$?
In order to address these questions, we calculate $\mathbf{Tr}\rho$ as a function of $\theta$ and $\phi$ after a particular number of iterations via Eq.~(\ref{eq:mat}).

To obtain the value of the damping parameter that will give $\delta_{min}$, we take the derivative of $\mathbf{Tr}\rho$ with respect to $\phi$ and set the resulting expression to zero. We then compare the behavior of the obtained damping parameter $(\phi_{opt})$ with that of $\phi_{crit}$ by plotting them both as a function of $\theta$.

\subsection{First iteration}
\noindent Replace the trace matrix of Eq.~\eqref{eq:mat} by $[\sin\theta\;\cos\theta\;1]^{T}$. We will get a column matrix with $\mathbf{Tr}\rho X$, $\mathbf{Tr}\rho Y$ and $\mathbf{Tr}\rho$ as the first, second and third entry of the final column matrix. This process will yield $\mathbf{Tr}\rho$ for the first iteration
\begin{equation}
	\mathbf{Tr}\rho(\theta, x)=
	\left(-\frac{1}{2}\cos\theta+\frac{1}{2}\right)x^2
	+\left(\frac{1}{2}\cos\theta+\frac{1}{2}\right),
\end{equation}
where $x=\cos\phi$. For a given $\theta$, $\mathbf{Tr}\rho$ is a quadratic function of $\cos\phi$. Since $0\leq\theta\leq\pi/2$ and $0\leq\phi\leq\pi/2$, then the minimum of $\mathbf{Tr}\rho$ is $\cos^{2}(\theta/2)$ which is at $\phi=\pi/2$. The $\delta$ that we must choose should not be less than the initial error probability, $\cos^{2}(\theta/2)$. Thus, for the first iteration, the optimized damping parameter is $\pi/2$ and is independent of the number of target states described by $\theta$. 

\subsection{Second iteration}
\begin{figure}[!hbp]  
\begin{center}
\mbox{\subfigure[$1^{st}$ iteration]{\includegraphics[trim = 0mm 0mm 0mm 170mm, clip, height=1.45in, width=0.5\columnwidth]{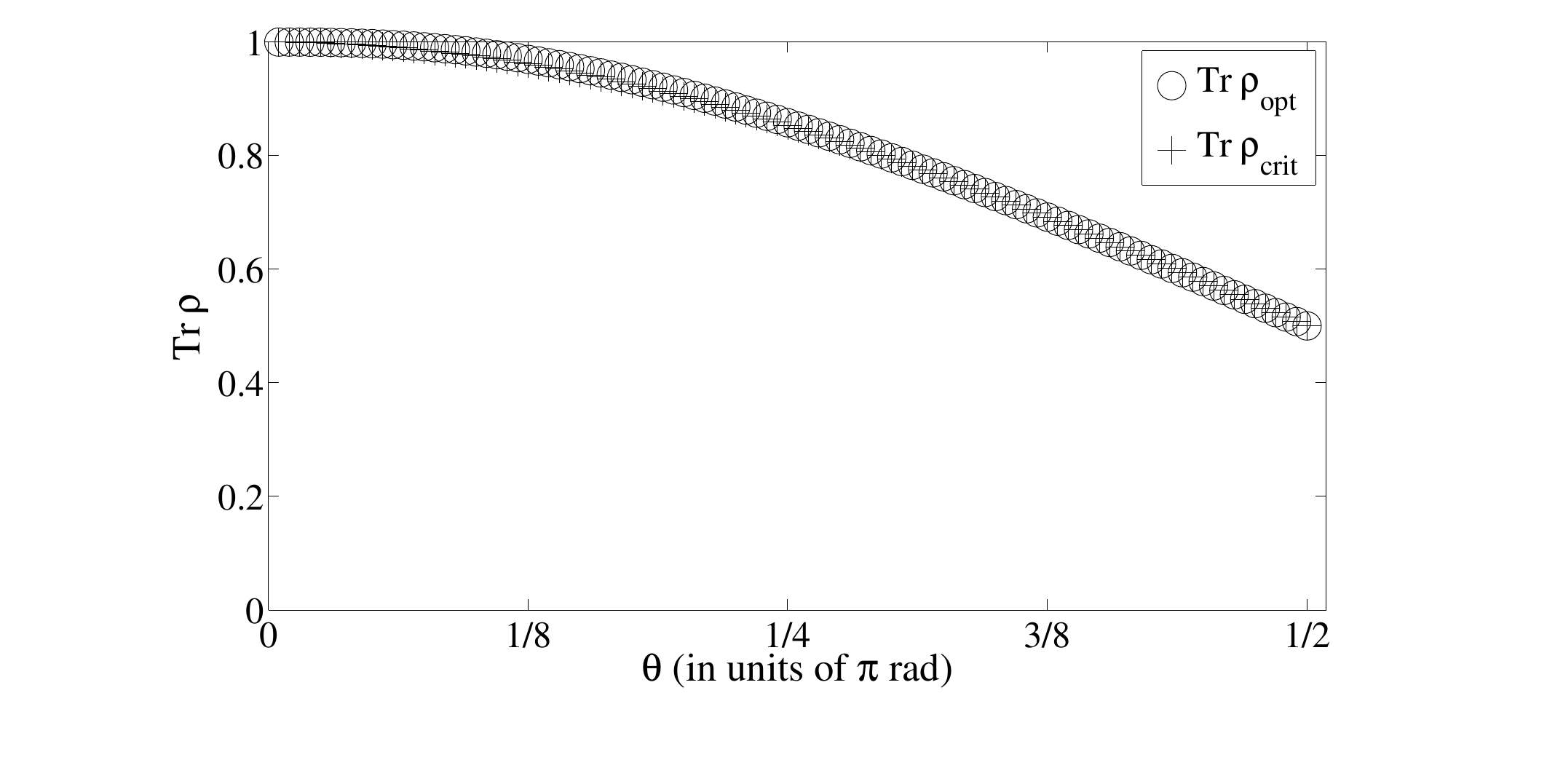}}\quad
\subfigure[$2^{nd}$ iteration]{\includegraphics[trim = 0mm 0mm 0mm 170mm, clip, height=1.5in, width=0.5\columnwidth]{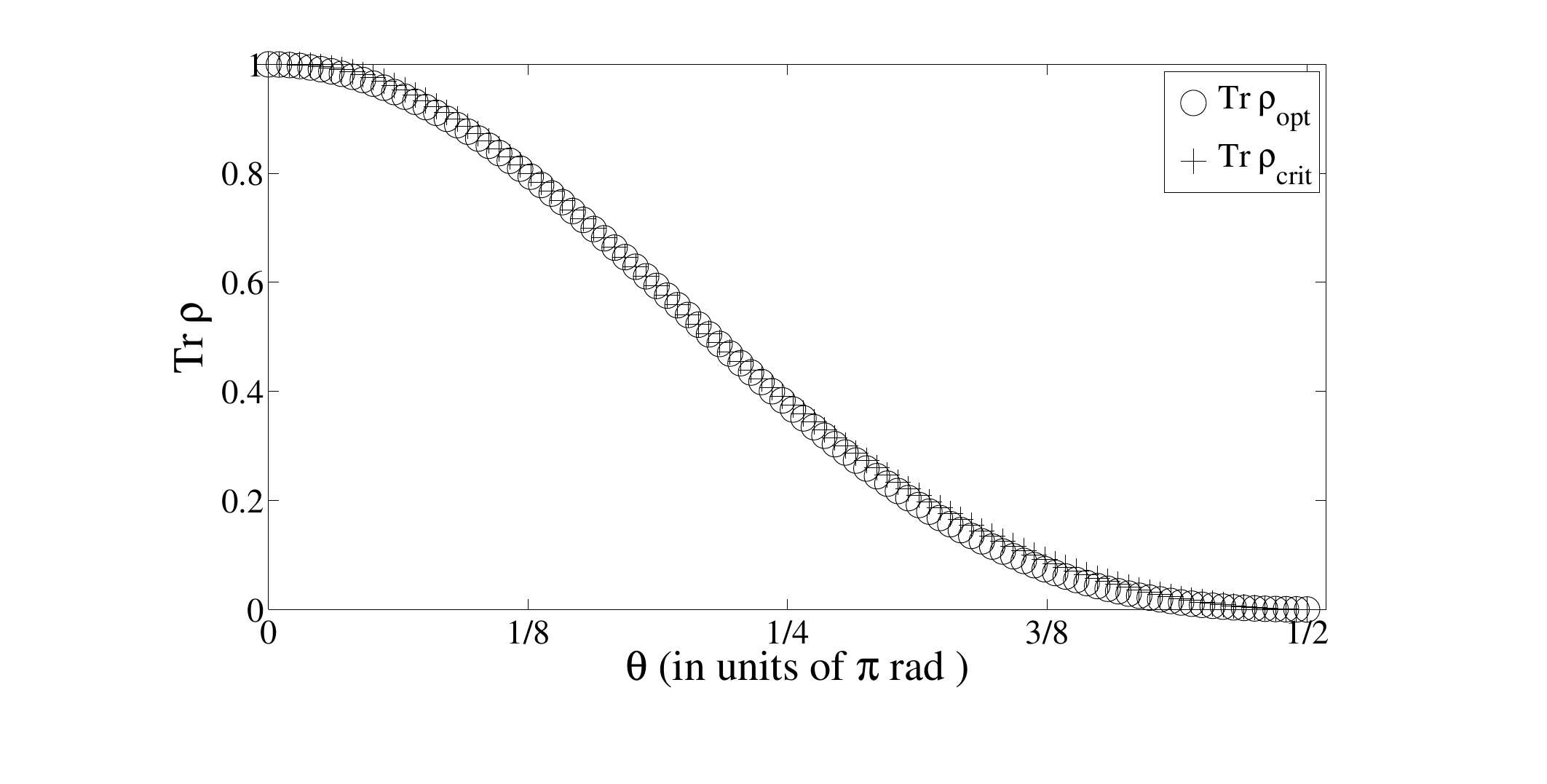} }}
\caption{$\mathbf{Tr}\rho(\phi_{opt}, \theta)$ and $\mathbf{Tr}\rho(\phi_{crit}, \theta)$ of the $1^{st}$ and the $2^{nd}$ iterations}
\label{fig:TrP1and2}
\end{center}
\end{figure}

\noindent
For the second iteration, $\mathbf{Tr}\rho$ is a quartic function of $x$ which implies that we have to solve the real root of a cubic polynomial in $x$ in order to obtain the extremum of $\mathbf{Tr}\rho$,
\begin{equation}\label{eq:TrP2nd}
		\begin{split}
			\mathbf{Tr}\rho(\theta, x)=	
 	 		\left(-\frac{1}{2}\cos^{3}\theta
 	 		+\frac{1}{2}\cos^{2}\theta\right)x^4
			+\left(-\cos^{3}\theta+\cos\theta\right)x^3\\
			+\left(-\cos^{2}\theta+1\right)x^2
			+\left(\cos^{3}\theta-\cos\theta\right)x
			+\left(\frac{1}{2}\cos^{3}\theta
			+\frac{1}{2}\cos^{2}\theta\right).
		\end{split}		
\end{equation}
The real root can still be obtained analytically and, as expected, it is no longer independent of $\theta$, 
\begin{equation}\label{eq:xopt2nd}
    \begin{split} 
      x_{opt}=
      -\frac{2}{3}\left(\frac{1+\cos\theta}{\cos\theta}\right)
      +\frac{[(1+\cos\theta)[(2-\cos\theta)^2
      +\sqrt{f(\cos\theta)}]]^{1/3}}{3\cos\theta}\\
      +\frac{4}{3}\left(\frac{1+\cos\theta}
      {[(1+\cos\theta)[(2-\cos\theta)^2
      +\sqrt{f(\cos\theta)}]]^{1/3}}\right),
    \end{split}
\end{equation}
where $f(\cos\theta)=-63\cos^4\theta-72\cos^3\theta+24\cos^3\theta-32\cos\theta+16$. Using Eq.~(\ref{eq:xopt2nd}) into Eq.~(\ref{eq:TrP2nd}) will give us $\delta_{min}$. Figure \ref{fig:TrP1and2} shows the plot of $\mathbf{Tr}\rho$ as a function of $\theta$ for the first and second iteration. There is a slight difference between the $\mathbf{Tr}\rho$ of the critical damped search and the $\mathbf{Tr}\rho$ of the optimum damped search for the first and second iteration.

The analytical calculations of $x_{opt}$ is no longer possible for higher iterations, i.e. for $n>2$, because generally there is no explicit formula for the real root of a $2n-1$ degree polynomial. So we resort to numerical calculations of the real root of a $2n-1$ degree polynomial in $x$ with $x\in[0,1]$. 

\section{Numerical calculations and results}
\label{sec:results}
\noindent

\begin{figure}[!htp]
\centering
\includegraphics[trim = 0mm 0mm 0mm 160mm, clip, height=2in, width=4in]{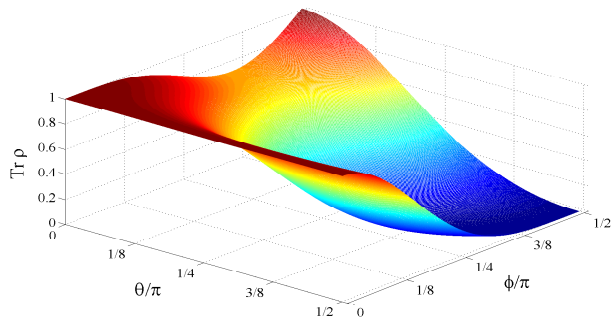}
\caption{The probability that the spin will not flip, $\mathbf{Tr}\rho$, as a function of $\theta$ and $\phi$ for the $3^{rd}$ iteration.}
\label{fig:3D_3rd}
\end{figure}

Figure~\ref{fig:3D_3rd} is a sample plot of $\mathbf{Tr}\rho$ as a function of $\theta$ and $\phi$ for the $3^{rd}$ iteration, showing that for a particular $\theta$ there exist a unique minimum $\mathbf{Tr}\rho$. The goal of this section is to obtain the value of $\phi$ that will give the minimum $\mathbf{Tr}\rho$ for a given $\theta$.

The calculation of $\phi_{opt}$ for $n>2$ iterations is carried out numerically as described in the following. The explicit forms of $\mathbf{Tr}\rho$ as a function of $\phi$ and $\theta$ for different iterations are calculated by taking the $n^{th}$ power of Eq. \eqref{eq:mat} and choosing the third entry after the resulting square matrix is pre-multiplied by $[\mathbf{Tr}(\rho X)\;\mathbf{Tr}(\rho Z)\; \mathbf{Tr}(\rho)]^{T}=[\sin\theta\;\cos\theta\;1]^{T}$. The resulting $\mathbf{Tr}\rho$ is differentiated with respect to $\cos\phi$ and is equated to zero. This will allow us to find the optimum value of $\phi$ that will give the minimum possible $\mathbf{Tr}\rho$. Only the root that is real and lies in the interval $[0,1]$ is chosen.
The behavior of $\phi_{opt}$ in each iteration is analyzed by plotting it as a function of $\theta$. This method is limited in a sense that the result will depend largely on the resolution of the grid. Nevertheless, this will provide a useful visual representation of $\phi_{opt}$ behavior for different $\theta$.

\begin{figure}[!htp]
	\begin{center}
\mbox{\subfigure[$4^{th}$ iteration]{\includegraphics[trim = 0mm 0mm 0mm 160mm, clip, height=1.3in, width=0.5\columnwidth]{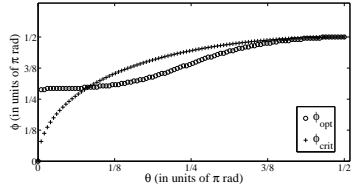}\label{fig:phia}}\quad
   	  \subfigure[$10^{th}$ iteration]{\includegraphics[trim = 0mm 0mm 0mm 160mm, clip, height=1.3in, width=0.5\columnwidth]{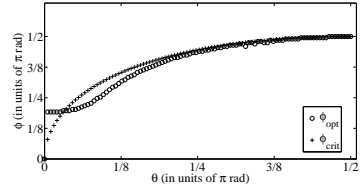}\label{fig:phib} }}
		\caption{Comparison of $\phi_{opt}$(o) and $\phi_{crit}$(+) as a function of $\theta$ for the $4^{th}$ and the $10^{th}$ iterations.}
		\label{fig:phi}
	\end{center}
\end{figure}
\begin{figure}[!htp]
\begin{center}
\mbox{\subfigure[$1^{st}$ to $4^{th}$ iterations]{\includegraphics[trim = 0mm 0mm 0mm 160mm, clip, height=1.3in, width=0.5\columnwidth]{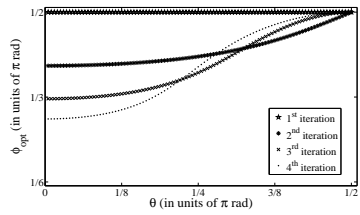}}\quad
   	  \subfigure[$5^{th}$, $7^{th}$ and $9^{th}$ iterations]{\includegraphics[trim = 0mm 0mm 0mm 160mm, clip, height=1.3in, width=0.5\columnwidth]{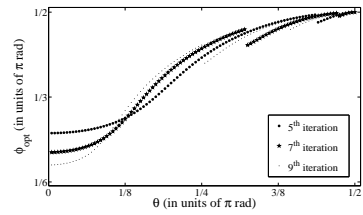}\label{fig:phi2b} }}
\caption{The changes in the range of $\phi_{opt}$ as the iteration is increased.}
\label{fig:phi2}
\end{center}
\end{figure}
\begin{figure}[!htp]  
	\begin{center}
	\mbox{\subfigure[$4^{th}$ iteration]{\includegraphics[trim = 0mm 0mm 0mm 160mm, clip, height=1.3in, width=0.5\columnwidth]{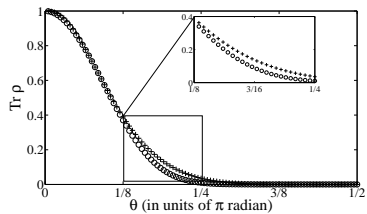}}\quad
	\subfigure[$10^{th}$ iteration]{\includegraphics[trim = 0mm 0mm 0mm 160mm, clip, height=1.3in, width=0.5\columnwidth]{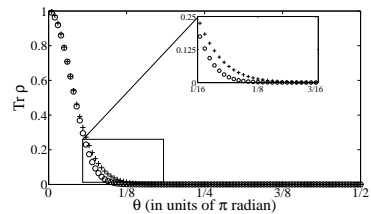} }}
		\caption{$\mathbf{Tr}\rho(\phi_{opt}, \theta)$ and $\mathbf{Tr}\rho(\phi_{crit}, \theta)$ of the $4^{th}$ and the $10^{th}$ iterations}
		\label{fig:TrP}
	\end{center}
\end{figure}
The plots of $\phi_{opt}$ as a function of $\theta$ are obtained for the third up to the tenth iterations. Only the fourth and tenth iterations of the optimum and the critical $\phi$ as function of $\theta$ plots are shown in Fig.~\ref{fig:phi} for convenience. The results show that for any number of iterations the critical damping is equal to the optimized damping parameter in the region near $\theta=\pi/2$, indicating that the critical damping is optimized for a target with a size comparable to that of the database. The leftmost part of the $\phi_{opt}$ curve will coincide eventually with the leftmost part of the $\phi_{crit}$ curve as we further increase the number of iterations. In Fig.~\ref{fig:phi2}, the range of the $\phi_{opt}$ is observed to increase as the number of iterations is increased. This implies that we do not have to use all the possible values of $\phi$ when optimizing the search for small number of iterations. Also, jumps begin to appear in the fifth iterations onwards as shown in Fig.~\ref{fig:phi2b}.

We have shown that the critical damping is not the same as the optimum damping that will give the smallest probability of failure. To continue with our analysis, we further investigate the difference of the resulting $\mathbf{Tr}\rho$ using $\phi_{opt}$ as compared to that of ${\phi_{crit}}$ for any $\theta$.

Using the numerically calculated $\phi_{crit}$, the $\mathbf{Tr}\rho$ are evaluated for different values of $\theta$ and for a different number of iterations. In Fig.~\ref{fig:TrP}, the values of $\mathbf{Tr}\rho$ are shown as a function of $\phi_{opt}$ and $\theta$ (denoted by $\circ$),  and as a function of $\phi_{crit}$ and $\theta$ (denoted by $+$). The values overlap for $M\sim N/2$ or $\theta\sim\pi/2$ but differ for small values of $\theta$. Notice that the interval of $\theta$, where the difference in $\mathbf{Tr}\rho$ is evident, goes to smaller values, i.e. smaller values of the target items, as the number of iterations is increased. It turns out that the critically damped quantum search becomes less optimized for larger number of iterations if the number of target states is small.

\begin{figure}[!htp]
	\begin{center}
\mbox{\subfigure[$4^{th}$ iteration]{\includegraphics[trim = 0mm 0mm 0mm 160mm, clip, height=1.3in, width=0.5\columnwidth]{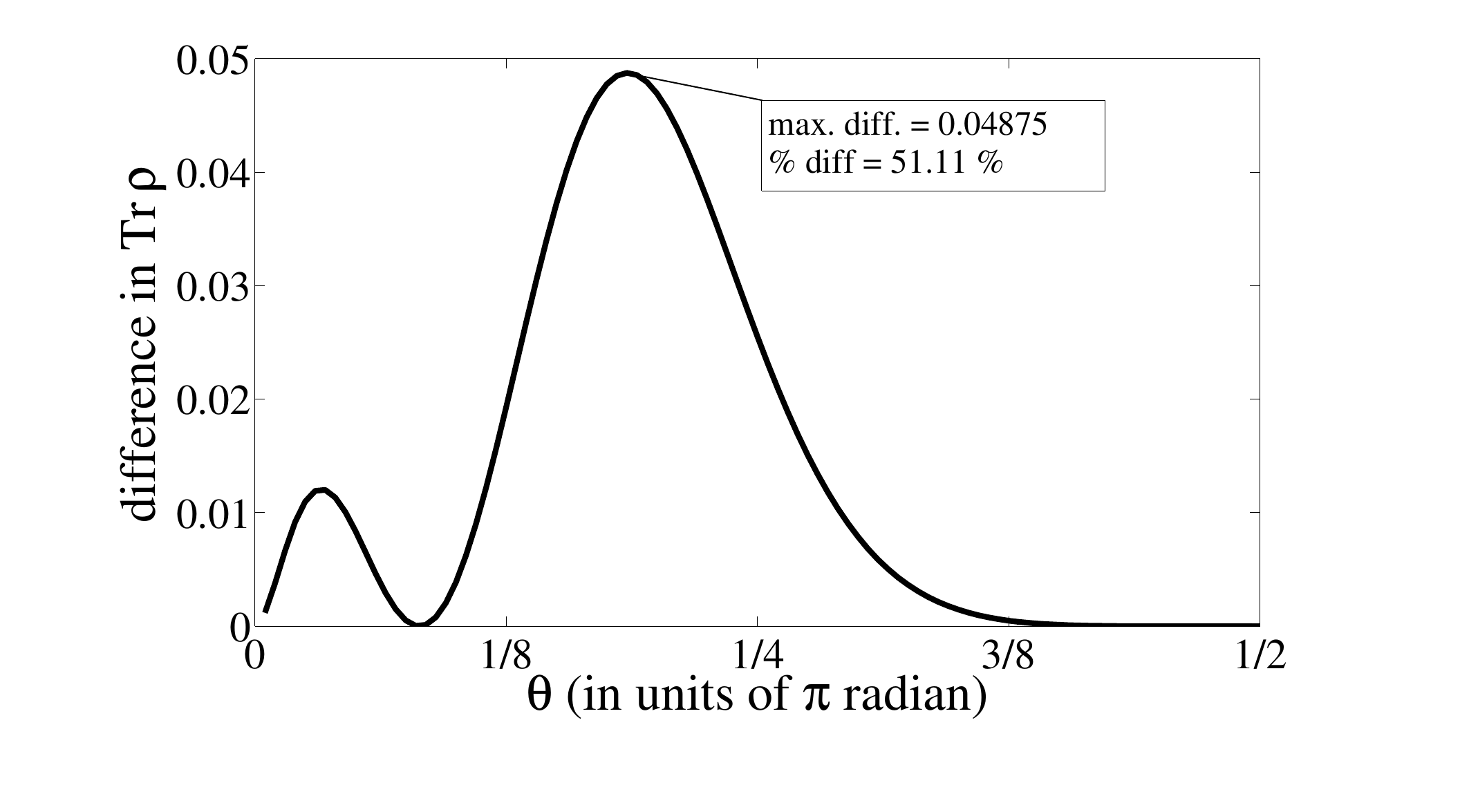}}\quad
   	  \subfigure[$10^{th}$ iteration]{\includegraphics[trim = 0mm 0mm 0mm 160mm, clip, height=1.3in, width=0.5\columnwidth]{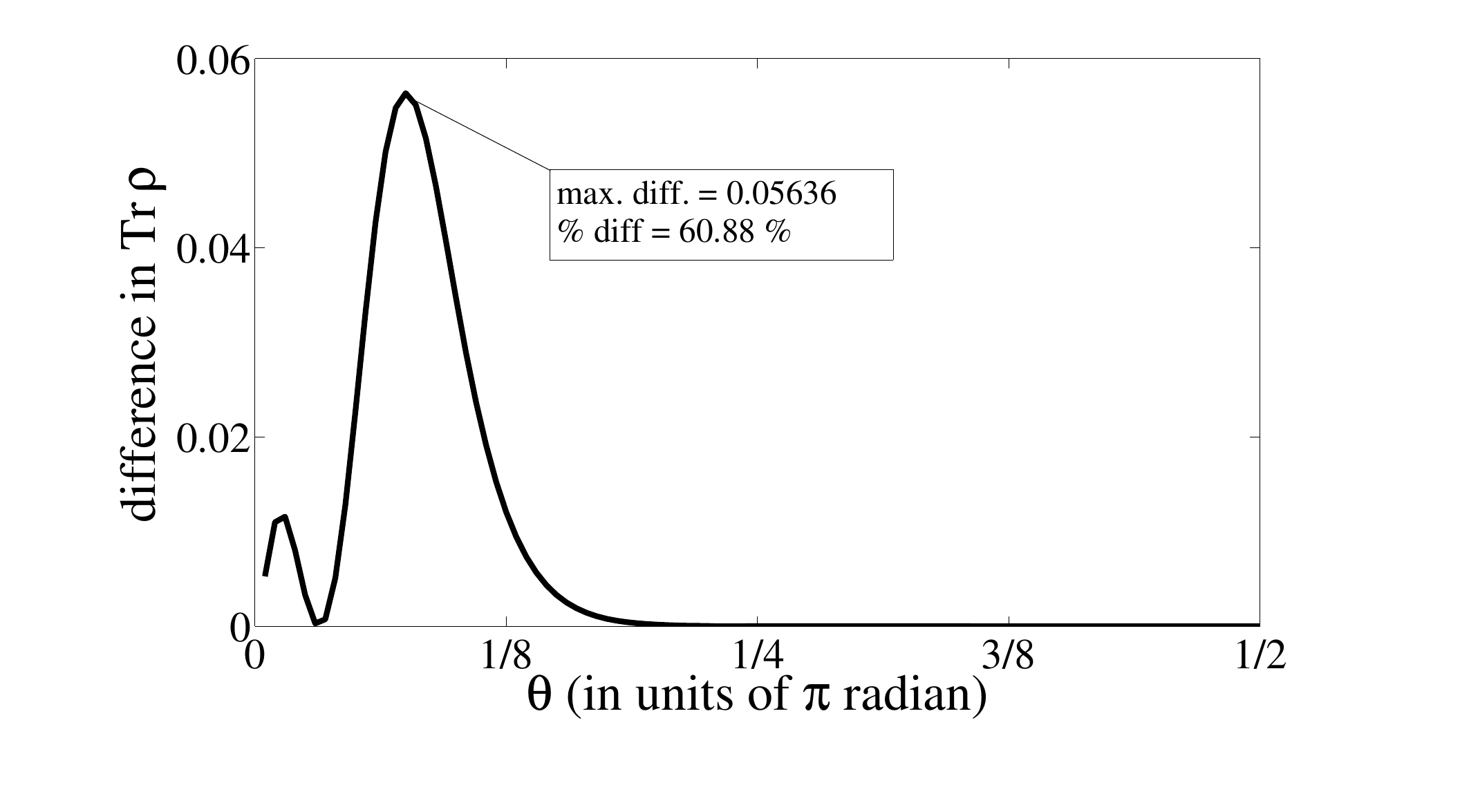} }}
		\caption{The plot of the difference in the probability between the critical damping parameter and the optimum damping parameter. 
		The inset also shows the corresponding percent difference of the peak.}
		\label{fig:diff}
	\end{center}
\end{figure}

\begin{table}[!htp]
\centering	
\caption{The maximum difference and the corresponding percentage difference between Tr$\rho$ of the optimized and the critical damping parameters.}
\begin{tabular}{c||cc}
\hline\hline
Iteration &difference	&\% difference \\
\hline
1st		&0.0077	&0.8105	\\
\hline
2nd	    &0.0194    &15.88	\\
\hline
3rd	    &0.0415    &43.52	\\
\hline
4th     &0.0487    &51.11	\\
\hline
5th  	&0.0520   &57.54	\\
\hline
6th	    &0.0538   &58.93	\\
\hline
7th	    &0.0547   &53.78	\\
\hline	
8th	    &0.0555   &64.53	\\
\hline
9th     &0.0559   &67.11	\\
\hline
10th    &0.0564   &60.88	\\
\hline\hline
\end{tabular}
\label{tab:8spinsVS}
\end{table}

In Fig.~\ref{fig:diff}, the difference in the probabilities $\mathbf{Tr}\rho(\phi_{opt}, \theta)$ and $\mathbf{Tr}\rho(\phi_{crit}, \theta)$ increases as the number of iterations is increased. The percent difference of the corresponding maximum difference are also shown in the plot. Table ~\ref{tab:8spinsVS} summarizes the maximum difference in the probabilities in each iteration. We observed that the percent difference with respect to the order of $\mathbf{Tr}\rho(\phi_{crit})$ of the maximum difference with $\mathbf{Tr}\rho(\phi_{opt})$ is more than 50\% from the fourth iteration onwards.

\section{Summary and Conclusion}
\label{sec:summary}
\noindent
An investigation was made on the damped quantum search by appealing directly to the global minimum of the failure probability. We have found that the critical damping parameter is generally not equal to the optimum damping parameter. There exists some interval in $\theta$ where the critically damped quantum search is evidently not optimized. In addition, the resulting failure probability is smaller compared to that of the critical damping parameter. However, such differences vanish readily as the number of target states approaches that of the database. Using the result of the optimum quantum search, we could develop an alternative way of calculating the damping parameter by solving a certain high order polynomial depending on the number of iterations. 
\section*{Acknowledgement}
\noindent
N.I. acknowledges support from the Department of Science and Technology SEI-ASTHRDP.




\begin{thebibliography}{100}
\bibitem{grover0}
L.K. Grover (1997),
{\it Quantum Mechanics Helps in Searching for a Needle in a Haystack},
Phys. Rev. Lett. 79 (2), pp 325-328.

\bibitem{Bennett}
C.H. Bennett, E. Bernstein, G. Brassard, and U. Vazirani (1997),
{\it Strength and weaknesses of quantum computing},
SIAM J. Comput. 26 (5), pp 1510-1523

\bibitem{Boyer}
M. Boyer, G. Brassard, P. Hoyer, and A. Tapp (1998),
{\it Tight bounds on quantum searching},
.Fortsch. Phys. - Prog. Phys., 46 (4-5), pp 493-505.

\bibitem{Liu}
Y. Liu and G. Koehler (2010),
{\it Using modifications to Grover's Search algorithm for quantum global optimization},
Eur. J. Oper. Res. 207 (2), pp 620-632.

\bibitem{Aaronson}
S. Aaronson and A. Ambainis (2005),
{\it Quantum search of spatial regions},
Theory of Computing (1), pp 47-79.

\bibitem{Cafaro}
C. Cafaro and S. Mancini (2012),
{\it On Grover's search algorithm from a quantum information geometry viewpoint},
Phys. A 391 (4), pp 1610-1625.

\bibitem{Farhi}
E. Farhi and S. Gutmann (1998),
{\it Analog analogue of a digital quantum computation},
Phys. Rev. A 391 57 (4) pp 2403-2406.

\bibitem{Childs}
A. Childs, E. Deotto, E. Farhi and J. Goldstone (2002),
{\it Quantum search by measurement},
Phys. Rev. A 391 66 (3).

\bibitem{Nielsen} 
Nielsen M A and Chuang I L (2000), {\it Quantum Computation and Quantum Information}, 
Cambridge University Press (Cambridge, England).

\bibitem{Brylinsky} 
Brylinsky R and Chen G (2002), {\it Mathematics of Quantum Computation}, 
Chapman and Hall/CRC (Florida).

\bibitem{fix1}
L.K. Grover (2005),
{\it Fixed Point Quantum Search},
Phys. Rev. Lett. 95, 150501.

\bibitem{fix2}
T. Tulsi, L. K. Grover, and A. Patel (2006),
{\it A new algorithm for fixed point quantum search}
Quantum Inf. Comput. 6, 483.

\bibitem{Grover2} 
Grover L K, Patel A, Tulsi T (2006), {\it Quantum algorithms with fixed points: The case of database search}, quant-ph/0603132.


\bibitem{mizel}
A. Mizel (2009),
{\it Critically Damped Quantum Search},
Phys. Rev. Lett. 102 (15), 150501.


\end{thebibliography}
\end{document}